# Sensitive Magnetic Control of Ensemble Nuclear Spin Hyperpolarisation in Diamond

Hai-Jing Wang[1,2], Chang S. Shin[1,2], Claudia E. Avalos[1,2], Scott J. Seltzer[1,2], Dmitry Budker[3,4], Alexander Pines[1,2], and Vikram S. Bajaj[1,2,*]

[1]*Materials Sciences Division, Lawrence Berkeley National Laboratory, Berkeley, California 94720, USA*

[2]*Department of Chemistry and California Institute for Quantitative Biosciences, University of California, Berkeley, California 94720, USA*

[3]*Nuclear Science Division, Lawrence Berkeley National Laboratory, Berkeley, California 94720, USA*

[4]*Department of Physics, University of California, Berkeley, California 94720, USA*

**Dynamic nuclear polarisation, which transfers the spin polarisation of electrons to nuclei, is routinely applied to enhance the sensitivity of nuclear magnetic resonance; it is also critical in spintronics, particularly when spin hyperpolarisation can be produced and controlled optically or electrically. Here we show the complete polarisation of nuclei located near the optically-polarised nitrogen-vacancy (NV) centre in diamond; by approaching the ground-state level anti-crossing condition of the NV electron spins, $^{13}$C nuclei in the first-shell are polarised in a pattern that depends sensitively and sharply upon the magnetic field. Based on the anisotropy of the hyperfine coupling and of the optical polarisation mechanism, we predict and**

**observe a complete reversal of the nuclear spin polarisation with a few-mT change in the magnetic field. The demonstrated sensitive magnetic control of nuclear polarisation at room temperature will be useful for sensitivity-enhanced NMR, nuclear-based spintronics, and quantum computation in diamond.**

Several techniques based on spin polarisation transfer have been developed to enhance the magnetisation of nuclei; they are applied widely in nuclear magnetic resonance (NMR)-based probes of the chemistry and structure of proteins, materials, and organisms[1-4], in devices that exploit spin-dependent electron transport in solids (spintronics)[5,6], in neutron scattering for structural characterization[7,8], and in quantum information experiments for the preparation of pure initial states[9,10]. Dynamic nuclear polarisation (DNP), in particular, is a family of techniques that transfer the greater spin polarisation of electrons to nearby nuclei through hyperfine-mediated processes[11,12]. In a typical DNP experiment at high magnetic fields, stable paramagnetic centres (usually organic radicals) are introduced into an otherwise diamagnetic sample as a polarisation source. The equilibrium polarisation of exogenous radical electrons is sufficiently high to polarise nuclei because the gyromagnetic ratio of electrons is at least three orders of magnitude higher than that of nuclear spins[13-15]. In addition to the high magnetic field, the experiment requires cryogenic temperatures to attenuate competing spin-lattice relaxation processes and relies upon strong microwave irradiation near the electron precession frequency for polarisation transfer. In alternative incarnations of this experiment that have been less generally applied, the initial electron polarisation is produced optically via the spin-orbit coupling[16] in a suitable substrate; the subsequent nuclear polarisation can be controlled by electrical current in certain semiconductors and at cryogenic temperatures[17-19].

The negatively charged nitrogen-vacancy (NV) centre in diamond (see Fig. 1a) can be prepared in a near-perfect state of electron spin polarisation by optical irradiation. Unlike in other systems, the high magnetic purity of the lattice in which the defect is situated means that this can be accomplished even at room temperature, making it particularly attractive as a general source of polarisation for NMR experiments or as a substrate for spintronic devices[20,21]. Although the NV defect has been already been exploited in many applications, such as solid-state magnetometry[22-24] and in single-photon sources[25], the precise and non-inductive control over its lattice nuclear polarisation, as is possible electrically in certain semiconductors but not for insulating solids, would greatly enhance its applications in sensitivity-enhanced NMR and spintronics[26]. Further, tightly coupled and polarised nuclear spins may serve both as quantum bits in information processing experiments and as auxiliary quantum registers that enhance the fidelity of certain algorithms[27-29].

Rather than exploiting electrical or radiofrequency (RF) control, we show that the nuclear spin polarisation in diamond can be manipulated optically and magnetically in a regime where the Zeeman coupling of the magnetic field to the electron spins is nearly equal to and cancels the zero-field splitting. The initial electron polarisation of the NV results from the optical pumping: the laser preferentially pumps the electron spin to the $|0\rangle$ sublevel of the spin triplet ($^3A$) in the ground state and creates electron spin polarisation on the order of unity. This occurs via a spin-dependent intersystem crossing from the excited triplet state ($^3E$) to the singlet state, from which the electron spin almost exclusively decays to the $|0\rangle$ sublevel of the ground state[30]. The laser can be gated to switch the electron polarisation on and off, but the polarisation does not depend linearly on the laser intensity. Furthermore, in general the electronic polarisation is not accompanied by nuclear polarisation. If the hyperfine interaction dominates the total Hamiltonian of the NV-nuclear spin system, then the electrons and nuclei can exchange spin angular momentum through a simultaneous spin flip-flop process, enabling spin-polarisation transfer[20]. Such a condition can be established near the level anti-crossing in the excited state (ESLAC at ~51 mT, see Fig. 1b); near 100% polarisation has been observed for nuclei such as $^{14}$N or $^{15}$N adjacent to the vacancy (i.e., the nuclei of the nitrogen component of the NV centre), as well as proximal $^{13}$C. In these experiments, a rather broad dependence of $^{14}$N or $^{15}$N polarisation on magnetic field suggests an absence of a magnetic control mechanism near ESLAC. However, the hyperfine interaction with $^{15}$N or $^{14}$N is isotropic in the excited state[31,32], whereas that with $^{13}$C is in general anisotropic and varies with the lattice site in the ground state[33,34]. The anisotropic hyperfine interaction can lead to a sensitive magnetic field dependence of electron-nuclear mixing and thus $^{13}$C spin polarisation, but this has been obscured by the short lifetime of the excited state, and by the minimal electron-nuclear mixing in the ground state near the ESLAC condition due to the different zero-field splittings in the ground and excited states[20,21].

The level anti-crossing condition can also be achieved in the ground state (GSLAC or LAC for short at ~102.5 mT, see Fig. 1b)[35], where the long lifetime allows the electron-nuclear mixing to be probed in detail. Here we demonstrate a near 100% nuclear polarisation of an ensemble of the first-shell $^{13}$C ($^{13}$C$_a$, see Fig. 1a) coupled to NV near LAC. It is convenient to study $^{13}$C$_a$ because the strong hyperfine interaction allows the transitions associated with the NV-$^{13}$C$_a$ pair to be clearly resolved from those of NV without $^{13}$C$_a$. We investigated, experimentally and theoretically, the sensitive dependence of such nuclear polarisation on the magnetic field. Probed by RF excitation in continuous-wave (CW) optically detected magnetic resonance (ODMR) spectra, the frequency distribution and the relative intensities of observed transitions reveal only one level anti-crossing condition for the NV-$^{13}$C$_a$ pair, at a magnetic field slightly higher than that for NV without $^{13}$C$_a$. Near such level anti-crossing condition, the nuclear polarisation depends sensitively on the magnetic field: at the most sensitive region near 105 mT, the nuclear polarisation changes from "spin-up" to "spin-down" and back to "spin-up," with a field change of only a few mT. The $^{13}$C$_a$ spin polarisation serves as an illustrative example for $^{13}$C at other lattice sites with different hyperfine parameters. The polarisation and control mechanism should apply to all hyperfine-coupled $^{13}$C, although each class of $^{13}$C grouped by the hyperfine parameters has different nuclear polarisation at the same magnetic field.

## Results

**ODMR spectra.** The degree of electron-nuclear spin mixing is determined by the relative strength of the hyperfine interaction in the total Hamiltonian; this also determines the energy levels and the resulting transition frequencies between two mixed states. The relative intensities of different transitions indicate the relative populations of the various mixed states. Both the frequencies and the relative intensities of allowed transitions can be measured in ODMR spectra[20]. Figure 2 shows representative ODMR spectra of the NV ensemble near LAC (see Supplementary Fig. S1-S2 online for spectra at ambient magnetic field and near LAC, respectively). The smaller peaks that are resolved from the intense primary peak are assigned to the NV with $^{13}C_a$[32,34,36]. Because there are only three equivalent $C_a$ sites for each NV, and because of the low natural abundance of $^{13}C$ (1.1%), the peak intensities (i.e., integrated area above spectra) associated with NV-$^{13}C_a$ pairs are much smaller than the primary peaks originating from the greater fraction of NV centres without $^{13}C_a$. The spectra taken near LAC may have one, two, three or four peaks, and are thus different from the spectra taken at ambient or low (<10 mT) magnetic field, where the two peaks associated with NV-$^{13}C_a$ pairs are separated by ~130 MHz and asymmetrically located on either side of the primary peaks[36]. The number of observed peaks near LAC suggests that the electron-nuclear spins are in mixed states, because otherwise selection rules only allow two electron spin transitions to be optically observed, each of which is associated with one nuclear spin state[29].

**Theoretical calculations of energy levels and eigenstates.** A full understanding of the spectra and the electron-nuclear mixed states near LAC requires the NV spin Hamiltonian including its hyperfine interaction with $^{13}C_a$, given by[37,38]:

$$H = D \cdot S_z^2 + E\left(S_x^2 - S_y^2\right) + \gamma_e \mathbf{B} \cdot \mathbf{S} + \mathbf{S} \cdot \mathbf{A} \cdot \mathbf{I} - \gamma_{^{13}C} \mathbf{B} \cdot \mathbf{I}, \qquad (1)$$

where the axial ($D \sim 2870$ MHz) and transverse ($E$, a strain splitting particular to each diamond sample; for our sample, $E \sim 3.5$ MHz) zero-field splitting (ZFS) parameters are determined from the spectra at ambient magnetic field (see Supplementary Fig. S1 online), **S** is the electron spin of NV with the electron gyromagnetic ratio of $\gamma_e$ and with $z$ defined along the NV symmetry axis, **B** is the external magnetic field, **I** is the $^{13}C_a$ nuclear spin with the gyromagnetic ratio, $\gamma_{^{13}C}$, and **A** is the hyperfine-interaction tensor between NV and $^{13}C_a$, which is given by $A_\perp = 123$ MHz and $A_p = 205$ MHz with its principal axis ~74° relative to the symmetry axis of the NV as characterized by ESR[39]. This hyperfine tensor can also be quantized along the NV symmetry axis through a simple rotation[12]:

$$\mathbf{S} \cdot \mathbf{A} \cdot \mathbf{I} = 199 \times S_x I_x + 123 \times S_y I_y + 129 \times S_z I_z - 21 \times (S_z I_x + S_x I_z), \quad (2)$$

where all numbers are in units of MHz, the $x$ axis lies within the NV-$^{13}C$ plane, and $y$ is perpendicular to the NV-$^{13}C$ plane, forming orthogonal coordinates (see Supplementary Fig. S3 online). There are two important consequences when the magnetic field approaches the level anti-crossing condition: first, no terms in Eq. 2 can be truncated because the hyperfine interaction is the dominant term in the Hamiltonian[12,40]; second, each term in the hyperfine interaction, such as $S_x I_x$, $S_y I_y$ and $S_x I_z$, may cause an additional level anti-crossing condition. These are clearly shown in the energy levels (Fig. 3a) and eigenstates (Fig. 3b and Supplementary Fig. S4-S5 online) calculated by diagonalizing the Hamiltonian. The mixed states, namely $|\alpha\rangle$, $|\beta\rangle$, $|\chi\rangle$, and $|\delta\rangle$, are defined as in Fig. 3a. The crossings of the eigenstate projections as shown in Fig. 3b are the signatures of the level anti-crossing conditions[37,39], among which the near-crossing between $|\alpha\rangle$ and $|\chi\rangle$ at ~100 mT (LAC-) and the crossing between $|\beta\rangle$ and $|\chi\rangle$ at ~105 mT (LAC+) are found at the lowest and highest magnetic field, respectively, near LAC at ~102.5 mT.

**The observed level anti-crossing condition.** It has been established that the NV electron spin is polarized to the $|0\rangle_z$ state under laser irradiation[30,41]. However, it is unclear how the spin population is distributed among the four electron-nuclear mixed states when the two electronic states are split near LAC by the large hyperfine interaction. The observed ODMR spectra could provide such detailed information, provided that each observed peak can be assigned to a pair of eigenstates connected by the transition. Figure 4a shows a comparison of the observed transition frequencies with those calculated from the energy levels as shown in Fig. 3a. The excellent agreement confirms that the low-intensity peaks originate from the NV-$^{13}$C pairs and enables each observed transition to be unambiguously assigned to a pair of eigenstates.

Based on the distribution of the observed transitions over ~30 mT range near LAC, only one level anti-crossing condition is observed for NV-$^{13}$C pairs, at LAC+, even though each hyperfine coupling term can potentially introduce an additional level anti-crossing condition. The first signature is the disappearance of the $|\alpha\rangle \leftrightarrow |\beta\rangle$ transition and the appearance of the $|\alpha\rangle \leftrightarrow |\chi\rangle$ and $|\alpha\rangle \leftrightarrow |\delta\rangle$ transitions as $B_z$ crosses LAC+ (Fig. 4). Without the hyperfine interaction, the transition frequency changes continuously when the energy of the polarized state $|0\rangle$ becomes greater than that of the $|-1\rangle$ state as $B_z$ crosses LAC (Fig. 1b). This was previously identified by the inverted phase of the electronic paramagnetic resonance (EPR) signal when sweeping the magnetic field across

LAC[42]. Due to the anisotropic hyperfine interaction, however, a frequency change will be anticipated when the polarized state goes from the lower to higher energy levels at the level anti-crossing condition, which is exactly what we observed at LAC+ (see Fig. 4a). The second signature is the disappearance of the $|\alpha\rangle \leftrightarrow |\beta\rangle$ transition between 104 mT and 105 mT (open black squares in Fig. 4), because the population in the $|\beta\rangle$ state becomes small, which was also shown by ESR[42]. Similar phenomena are not observed at LAC-. These observations suggest that the mixed state with the greatest polarisation or population changes from $|\beta\rangle$ to $|\chi\rangle$ at LAC+, but not from $|\alpha\rangle$ to $|\chi\rangle$ at LAC-.

These phenomena are the inherent consequences of the spin polarisation mechanism of the NV. With the NV electron spins polarised in the $|0\rangle_z$ state by optical pumping, the only electron spin transition contributing to the polarisation transfer to nuclei is from $|0\rangle_z$ to $|-1\rangle_z$ via the simultaneous spin flip-flop induced by $S\_I_+$. This hyperfine element causes the accumulation of the nuclear spin population in the $|\uparrow\rangle_z$ state. Therefore the continuous laser irradiation and polarisation transfer tend to populate the $|0,\uparrow\rangle_z$ state. This explains why only LAC+ is observed, occurring at the magnetic field where the two mixed states with the largest $|0,\uparrow\rangle_z$ component, namely $|\beta\rangle$ and $|\chi\rangle$, meet the level anti-crossing condition (see Fig. 3b).

**The electron spin polarisation.** With each transition assigned to the mixed states and the effective level anti-crossing conditions identified, the relative peak intensities within the same spectrum reveal the spin polarisation below and above LAC+. The spin polarisation among the four mixed states is illustrated in Fig. 3a and explained as follows: From ~88 mT to ~96 mT, $I_{|\alpha\rangle\leftrightarrow|\delta\rangle}$ (the intensity of the integrated area of the peak assigned to the $|\alpha\rangle \leftrightarrow |\delta\rangle$ transition) decreases while $I_{|\beta\rangle\leftrightarrow|\delta\rangle}$ increases (see Fig. 2), suggesting that the spin population of $|\alpha\rangle$ is gradually pumped into $|\beta\rangle$. From ~96 mT to ~105 mT, the $|\alpha\rangle \leftrightarrow |\delta\rangle$ transition disappears and all transitions are associated with the $|\beta\rangle$ state (see Fig. 2b and Fig. S2a), which suggests that the spin population is nearly polarised to the $|\beta\rangle$ state.

Above 106 mT, the majority of the spin population should change from $|\beta\rangle$ to $|\chi\rangle$ due to the level anti-crossing at LAC+. A relatively small $I_{|\alpha\rangle\leftrightarrow|\delta\rangle}$ compared to $I_{|\alpha\rangle\leftrightarrow|\chi\rangle}$ and $I_{|\beta\rangle\leftrightarrow|\chi\rangle}$ suggests that the majority of the spins are in the $|\chi\rangle$ state, with only a small fraction in the $|\delta\rangle$ state (see Fig. S2c). An exception is found within a very narrow region of 105-106 mT, where $I_{|\alpha\rangle\leftrightarrow|\chi\rangle}$ is smaller than $I_{|\alpha\rangle\leftrightarrow|\delta\rangle}$ (see Fig. S2b). The spin population in $|\delta\rangle$ is slightly larger than that in $|\chi\rangle$ due to the tendency of the optical pumping mechanism to maximise $|0\rangle_z$; the $|\delta\rangle$ state has the largest projection onto $|0\rangle_z$ (or $|\langle 0|\delta\rangle|^2$) within this field range (see Fig. 3b). The $|\chi\rangle \leftrightarrow |\delta\rangle$ transition was not observed in ODMR spectra because this transition is not associated with significant change in the state of the electron spin, and thus there is little optical contrast.

Our tabulation of observed transition frequencies demonstrates how spin polarisation is affected by the anisotropic hyperfine interaction near the level anti-crossing. The detailed description of the spin polarisation and its dependence on the magnetic field provide insights into the mechanism of electron and nuclear spin polarisation using the NV centre[30].

**Nuclear polarisation and its dependence on magnetic field.** Both the polarisation of the electron-nuclear mixed states and their projections onto each nuclear spin state depend on the magnetic field, suggesting that the nuclear polarisation should as well. With the eigenstates as previously calculated (see Fig. 3b and Supplementary Fig. S4-S5 online), the spin population of each electron-nuclear mixed state is needed for a precise determination of the nuclear polarisation. In principle, the spin population can be calculated from the peak intensities of all transitions between any two mixed states. However, some transitions associated with NV-$^{13}C_a$ pairs may be obscured by the main peak from NV centres without $^{13}C_a$ (see the grey area in Fig. 4a). Nevertheless, we can still estimate the nuclear polarisation under the assumption that the electron polarisation is near 100%.

The estimated nuclear polarisation ($P$) based on the level anti-crossing between the $|0\rangle$ and $|-1\rangle$ sublevels ($P_{|0\rangle\leftrightarrow|-1\rangle}$, red up triangles, see Methods for the estimation of nuclear polarisation) is shown in Fig. 4b as a function of the magnetic field. The nuclear polarisation increases as $B_z$ increases until ~96 mT because the normalized spin population in $|\beta\rangle$ ($p_\beta$) is much greater than that of $|\alpha\rangle$ ($p_\alpha$). Above LAC+, the spin distribution among $|\chi\rangle$ and $|\delta\rangle$ causes the polarisation to plateau around ~0.7 up to ~116 mT. The nuclear polarisation decreases in the vicinity of LAC+ because of the approximately equal projection onto nuclear spin-up and spin-down of both the $|\beta\rangle$ and $|\chi\rangle$ states ($|\langle\uparrow|\beta\rangle|^2 \approx |\langle\downarrow|\beta\rangle|^2$ and $|\langle\uparrow|\chi\rangle|^2 \approx |\langle\downarrow|\chi\rangle|^2$, see Fig. 3b and Supplementary Fig. S4-S5 online). At 105.2 mT, the polarisation is negative because $p_\delta$ is slightly larger than $p_\chi$, and $|\langle\downarrow|\delta\rangle|^2$ is larger than $|\langle\uparrow|\chi\rangle|^2$. These results demonstrate not only that the nuclei can be completely hyperpolarised, but also that the magnitude and direction of the polarisation depend very sensitively on the magnetic field near the LAC.

The nuclear polarisation is also probed directly via the $|0\rangle \leftrightarrow |+1\rangle$ transitions ($P_{|0\rangle\leftrightarrow|+1\rangle}$, blue down triangles in Fig. 4b, see Methods for the estimation of nuclear polarisation). Since the $|+1\rangle$ sublevel is well-separated from the $|0\rangle$ and $|-1\rangle$ sublevels near the LAC, both $|+1,\uparrow\rangle$ and $|+1,\downarrow\rangle$ are nearly pure states. The population of nuclear spin-up or spin-down is therefore directly proportional to the peak intensity of the transition leading to $|+1,\uparrow\rangle$ or $|+1,\downarrow\rangle$, respectively (see Supplementary Fig. S6 online). The $P_{|0\rangle\leftrightarrow|+1\rangle}$ shows similar sensitive dependence on the magnetic field as $P_{|0\rangle\leftrightarrow|-1\rangle}$, but has more nuclear spin-down than $P_{|0\rangle\leftrightarrow|-1\rangle}$. This is because when the $|0\rangle \rightarrow |+1\rangle$ transition is excited by the resonant RF, the $S_+I_-$ term also contributes to the polarisation transfer from electrons to nuclei, leading to more nuclear spin-down. The nuclear spin can be completely polarised to the spin-down state near LAC+. The nuclear polarisation measured via different electronic sublevels of NV thus confirms its sensitive dependence on the magnetic field and the assumptions of our theory.

**Discussion**

As we have demonstrated elsewhere[43], shaped optical irradiation can be used to produce a time- and space-varying pattern of polarisation up to the diffraction limit; here we demonstrate that the magnetic field can act as a complementary control parameter. The anisotropic hyperfine interaction of NV centres with the surrounding lattice carbons in diamond allows the magnetic field to be used as an active and sensitive control of the nuclear polarisation near 105 mT (LAC+). The effect of the hyperfine interaction is pronounced only at the near-degeneracy condition of electronic states, which is at the level anti-crossing for the NV centres in diamond. A similar nuclear polarisation mechanism has also been used in GaAs semiconductors at the fractional quantum Hall regime with a filling factor of $\nu = 2/3$ [44-46]. When the cyclotron energies of the first two Landau levels match the electron spin Zeeman energy at the filling factor of $\nu = 2/3$, the electronic states become nearly degenerate and the nuclear spin can be polarised by electrical current and measured by resistivity[46]. Unfortunately, cryogenic temperatures (<4.2 K) and high magnetic field (a few Tesla) are generally required for such nuclear polarisation in semiconductors. The advantages of the NV centre in diamond, by contrast, are that similar control over polarisation can be achieved at room temperature and low magnetic field (~0.1 T). The combination of optical pumping and initialization, precise magnetic control, and optically detected polarisation as demonstrated here are a complete set of non-inductive techniques that are potentially useful for nuclear-based spintronics in insulating or semiconducting diamonds[25,26]. Our experiments, demonstrating complete polarization of the $^{13}$C nuclei at ~0.1T and room temperature, together with the results of King et al.[47], demonstrating transport of polarization to bulk nuclei via spin diffusion at

~9.4T, suggest that the bulk polarization of $^{13}$C nuclei in our diamond sample approaches several percent. Thus, our method may be applied wherever hyperpolarized bulk $^{13}$C nuclei[47] are required, including in DNP-enhanced NMR experiments, in spintronic devices, and for NV-based devices including magnetometers and gyroscopes[48].

## Methods

**Materials and experiments**. The diamond used in our experiments contains ~2 ppm of negatively charged NV (labelled S9 in previous studies[23,49]). A continuous beam from a 532 nm laser with optical power of 2 mW is focused on the diamond by an objective lens with a numerical aperture of 0.7 to achieve optical power density of ~ 5 mW/$\mu$m$^2$. The fluorescence signal from NV is detected by an avalanche photo detector after passing a dichroic mirror and a long-pass filter[49]. The external magnetic field is applied with a permanent magnet aligned to one of the NV symmetry axes with a precision better than ~1°. The strength of the magnetic field is adjusted by the distance of the magnet from the laser focus point on the diamond. The magnetic field is determined precisely using the NV without proximal $^{13}$C$_a$ nuclei acting as a magnetometer; the central frequencies of the intense primary peaks in the ODMR spectra are directly related to the magnetic field[23]. A single copper wire loop arranged about the laser focus produces an oscillating RF magnetic field with Rabi nutation frequency ~350 kHz. The ODMR spectra are measured by recording the fluorescence intensity while the excitation frequency is scanned from 1 to 500 MHz, normalized to the fluorescence intensity without RF excitation.

**Estimation of nuclear polarisation.** Assuming that the electron polarisation is near 100%, almost all spin population resides either at $|\alpha\rangle$ or at $|\beta\rangle$ below LAC+, and either at $|\chi\rangle$ or at $|\delta\rangle$ above LAC+, i.e., the normalized spin population ($p_{|\psi\rangle}$) is approximately given by $p_\alpha + p_\beta = 1$ and $p_\chi = p_\delta = 0$ below LAC+. In addition, the ODMR peak intensity is also affected by the optical contrast between $|0\rangle$ and $|-1\rangle$ in the excited states. Below LAC+, the spin population can be related to the observed peak intensities of $I_{|\beta\rangle\leftrightarrow|\delta\rangle}$ and $I_{|\alpha\rangle\leftrightarrow|\delta\rangle}$ by considering the fluorescence contrast of the two states:

$$\frac{I_{|\beta\rangle\leftrightarrow|\delta\rangle}}{I_{|\alpha\rangle\leftrightarrow|\delta\rangle}} = \frac{p_\beta \times \left(|\langle 0|\beta\rangle|^2 - |\langle 0|\delta\rangle|^2\right)}{p_\alpha \times \left(|\langle 0|\alpha\rangle|^2 - |\langle 0|\delta\rangle|^2\right)}. \tag{3}$$

The nuclear polarisation, $P$, can be estimated based on the polarisation among the mixed states and their projections onto each nuclear spin state:

$$P_{|0\rangle\leftrightarrow|-1\rangle} = p_\beta \times \left(|\langle\uparrow|\beta\rangle|^2 - |\langle\downarrow|\beta\rangle|^2\right) + p_\alpha \times \left(|\langle\uparrow|\alpha\rangle|^2 - |\langle\downarrow|\alpha\rangle|^2\right). \tag{4}$$

The transitions associated with $|\chi\rangle$ are obscured by the main peak (the grey area in Fig. 4a), and we may disregard these transitions under the assumption that $p_\chi = p_\delta = 0$. A similar method is used to calculate $P$ above LAC+ by comparing $I_{|\alpha\rangle\leftrightarrow|\chi\rangle}$ with $I_{|\alpha\rangle\leftrightarrow|\delta\rangle}$, and by assuming that $p_\alpha = p_\beta = 0$ and $p_\chi + p_\delta = 1$.

The nuclear polarisation measured via the $|0\rangle \leftrightarrow |+1\rangle$ transitions, $P_{|0\rangle \leftrightarrow |+1\rangle}$, is calculated by directly comparing the intensities of the transitions to $|+1,\uparrow\rangle$ ($I_{|+1,\uparrow\rangle}$) and to $|+1,\downarrow\rangle$ ($I_{|+1,\downarrow\rangle}$):

$$P_{|0\rangle \leftrightarrow |+1\rangle} = \left(I_{|+1,\uparrow\rangle} - I_{|+1,\downarrow\rangle}\right) \Big/ \left(I_{|+1,\uparrow\rangle} + I_{|+1,\downarrow\rangle}\right). \tag{5}$$

## Acknowledgements

This work was supported by the Director, Office of Science, Office of Basic Energy Sciences, Materials Sciences and Engineering Division, of the U.S. Department of Energy under Contract No. DE-AC02-05CH11231. D.B. acknowledges salary support from IMOD and the AFOSR/DARPA QuSAR program. We thank Prof. Jeffrey Reimer for his helpful advice about our manuscript.

## Author contributions

H.J.W. and V.S.B. conceived the idea and planned the project, H.J.W., C.S.S., and C.E.A. constructed the experimental apparatus, H.J.W. and C.S.S. performed the experiments, H.J.W. and S.J.S. analysed the data, H.J.W., V.S.B., S.J.S. and C.S.S. wrote the manuscript, D.B., A.P. and V.S.B. oversaw the project, and all authors discussed the results and commented on the manuscript.

## Additional information

**Competing financial interests:** The authors declare no competing financial interests.

# Figure legends

**Figure 1 | Atomic structure and energy diagram of NV in diamond. (a)** Atomic representation of NV centre in diamond lattice with one of the first-shell carbon sites ($C_a$) occupied by $^{13}C$ (red). **(b)** The energy diagram of the ground state triplet ($^3A$, in solid lines) and the excited state triplet ($^3E$, dashed lines) versus the magnetic field. The level anti-crossing occurs at ~51 mT in the excited state (ESLAC, dashed circle) and at ~102.5 mT in the ground state (GSLAC, solid circle).

**Figure 2 | Representative ODMR spectra near LAC.** The spectra are acquired at $B_z \sim 90.8$ mT (a) and $B_z \sim 96.4$ mT (b). All peaks are fit to Gaussian lineshapes. The intense primary peak associated with NV centres without $^{13}C_a$ is used to determine the magnetic field as shown in light blue solid lines. The weak peaks associated with NV-$^{13}C_a$ pairs are colour-coded and labelled according to the related transitions determined in Fig. 4a. The peak labelled by a red * is an artefact whose second harmonic matches the frequency of the main peak.

**Figure 3 | Theoretical calculations of energy levels and eigenstates in the NV-$^{13}C_a$ system. (a)** Energy of the four mixing states near LAC. The LAC-, LAC, and LAC+ are labelled by vertical dashed lines. The spin symbols illustrate how the polarisation evolves at different magnetic field strength, given the dominant nuclear projection of each state. **(b)** Eigenstates are expressed by their projections onto the $\left| m_S = 0, m_I = \uparrow / \downarrow \right\rangle_z$ states of the NV-$^{13}C_a$ pair, where $m_S$ and $m_I$ are electron and nuclear quantum numbers, respectively. The LAC-, LAC, and LAC+ are labelled by circles.

**Figure 4 | Transition assignments for the estimation of nuclear polarisation. (a)** Assignment of the observed transitions (solid symbols) based on frequency in comparison with theoretical calculation (solid lines). Some allowed transitions related to the NV-$^{13}C_a$ pair may be obscured by the main peak from NV without $^{13}C_a$ (grey area). The disappearance of certain transitions near LAC+ is shown by open black squares. **(b)** Nuclear polarisation measured via different electronic sublevels of the NV. $P_{|0\rangle\leftrightarrow|-1\rangle}$ is shown in red up triangles, and $P_{|0\rangle\leftrightarrow|+1\rangle}$ is shown in blue down triangles. The solid lines serve as a guide for the eye.

**FIG. 1**

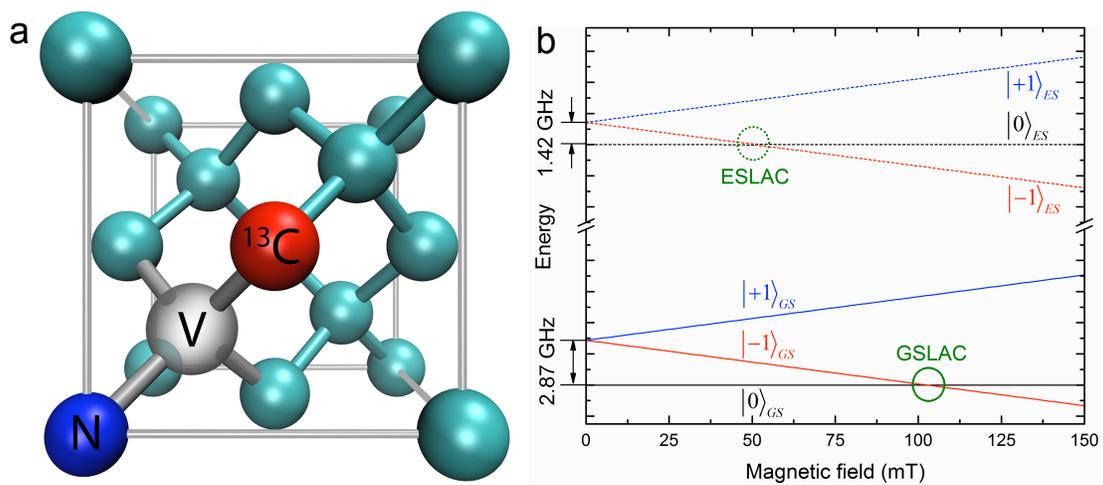

**FIG. 2**

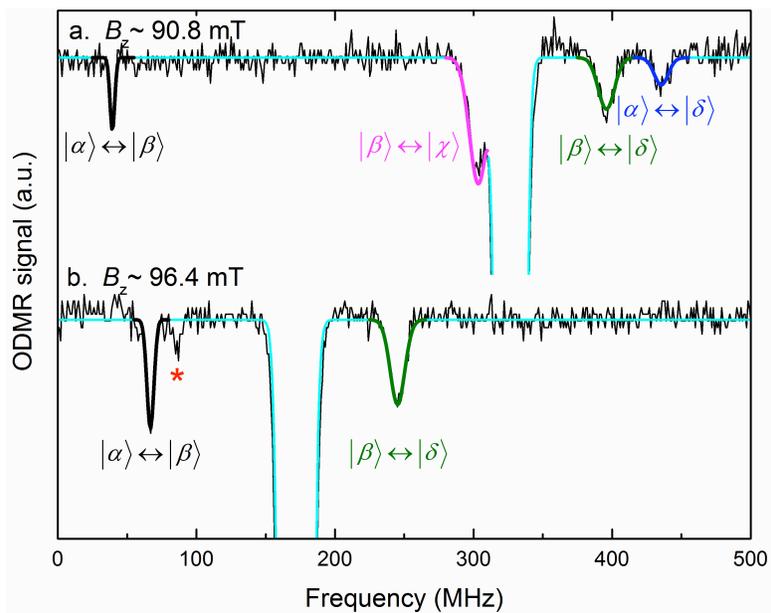

**FIG.3**

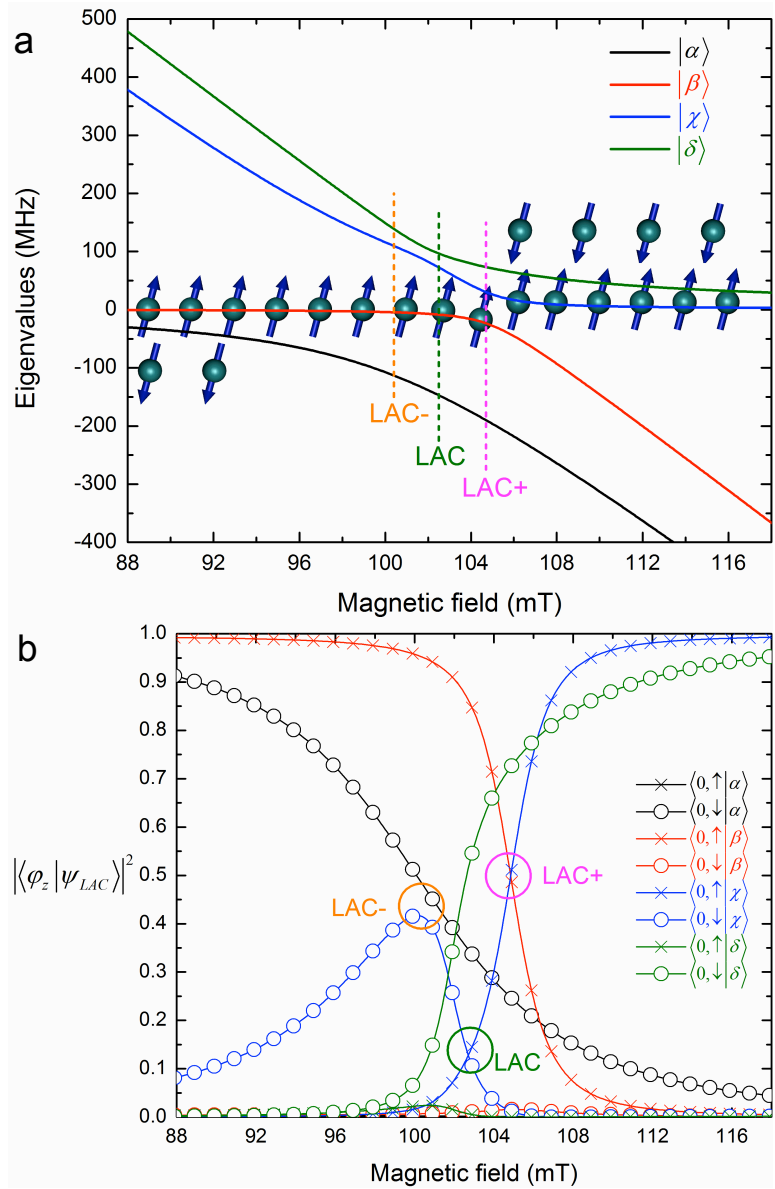

**FIG. 4**

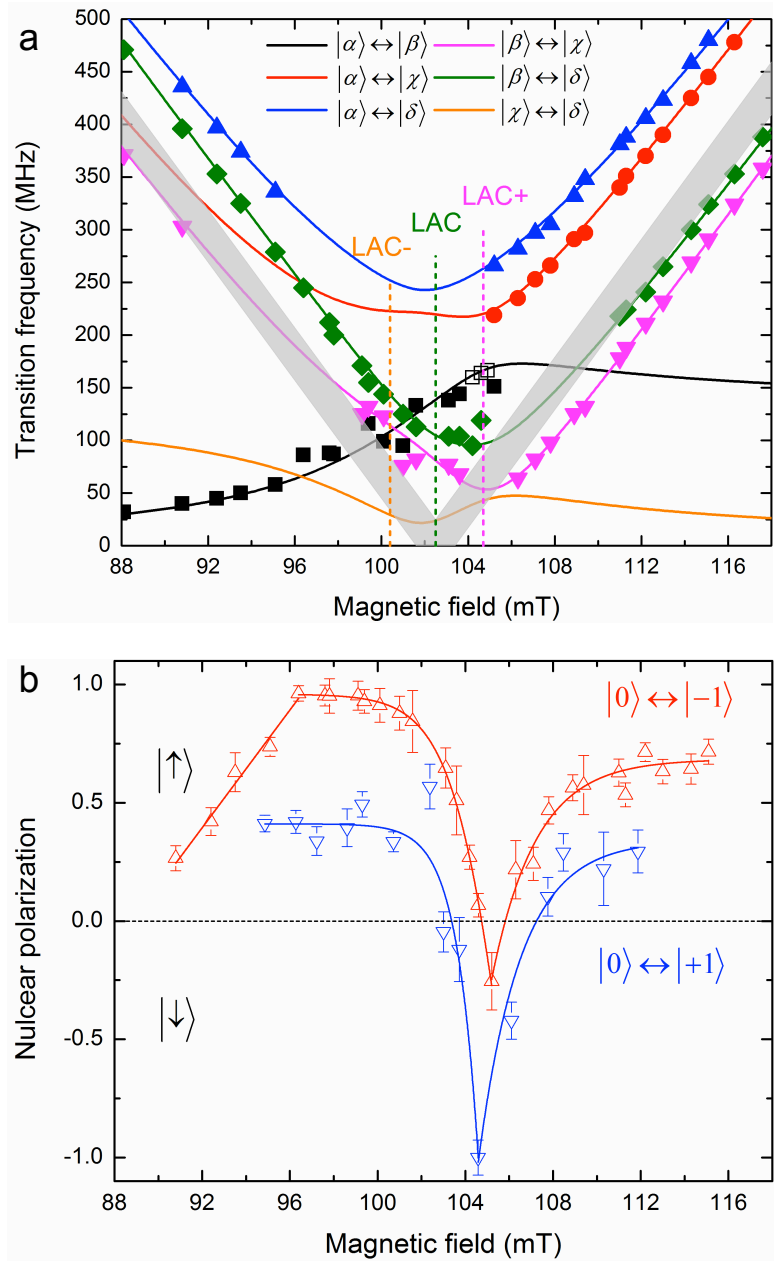

# Supplementary Informaiton

# Sensitive Magnetic Control of Ensemble Nuclear Spin Hyperpolarisation in Diamond

Hai-Jing Wang[1,2], Chang S. Shin[1,2], Claudia E. Avalos[1,2], Scott J. Seltzer[1,2], Dmitry Budker[3,4], Alexander Pines[1,2], and Vikram S. Bajaj[1,2,*]

[1]*Materials Sciences Division, Lawrence Berkeley National Laboratory, Berkeley, California 94720, USA*

[2]*Department of Chemistry and California Institute for Quantitative Biosciences, University of California, Berkeley, California 94720, USA*

[3]*Nuclear Science Division, Lawrence Berkeley National Laboratory, Berkeley, California 94720, USA*

[5]*Department of Physics, University of California, Berkeley, California 94720, USA*

**Supplementary Figures S1-S6**

**Supplementary Figure S1**

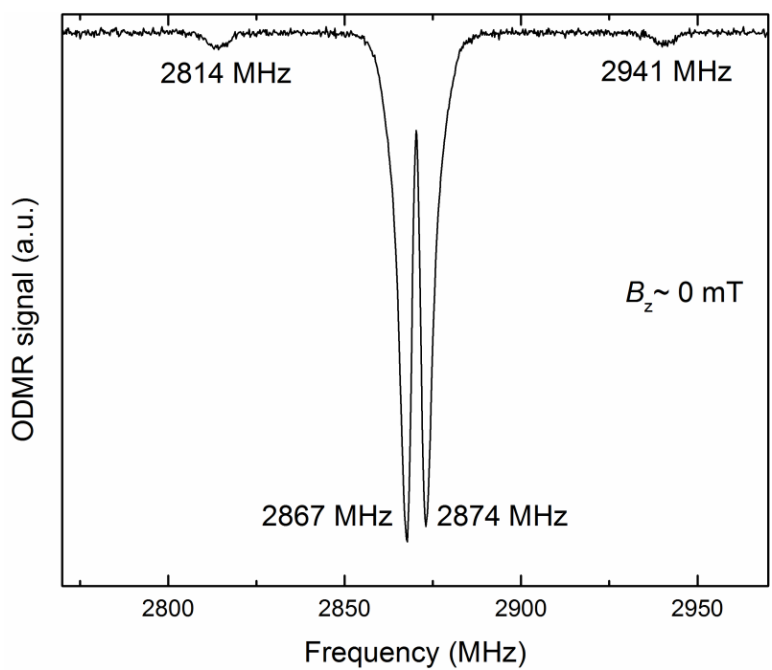

**Supplementary Figure S1 | ODMR spectrum of NV ensemble at ambient magnetic field ($B_z$~0 mT).** The observed transition associated with NV-$^{13}C_a$ pairs (relatively weak peaks at 2814 and 2941 MHz on either side of the main peaks) are in excellent agreement with the calculated transition frequencies (2813 MHz and 2941 MHz) based on theoretical calculation of the energy levels in the main text.

**Supplementary Figure S2**

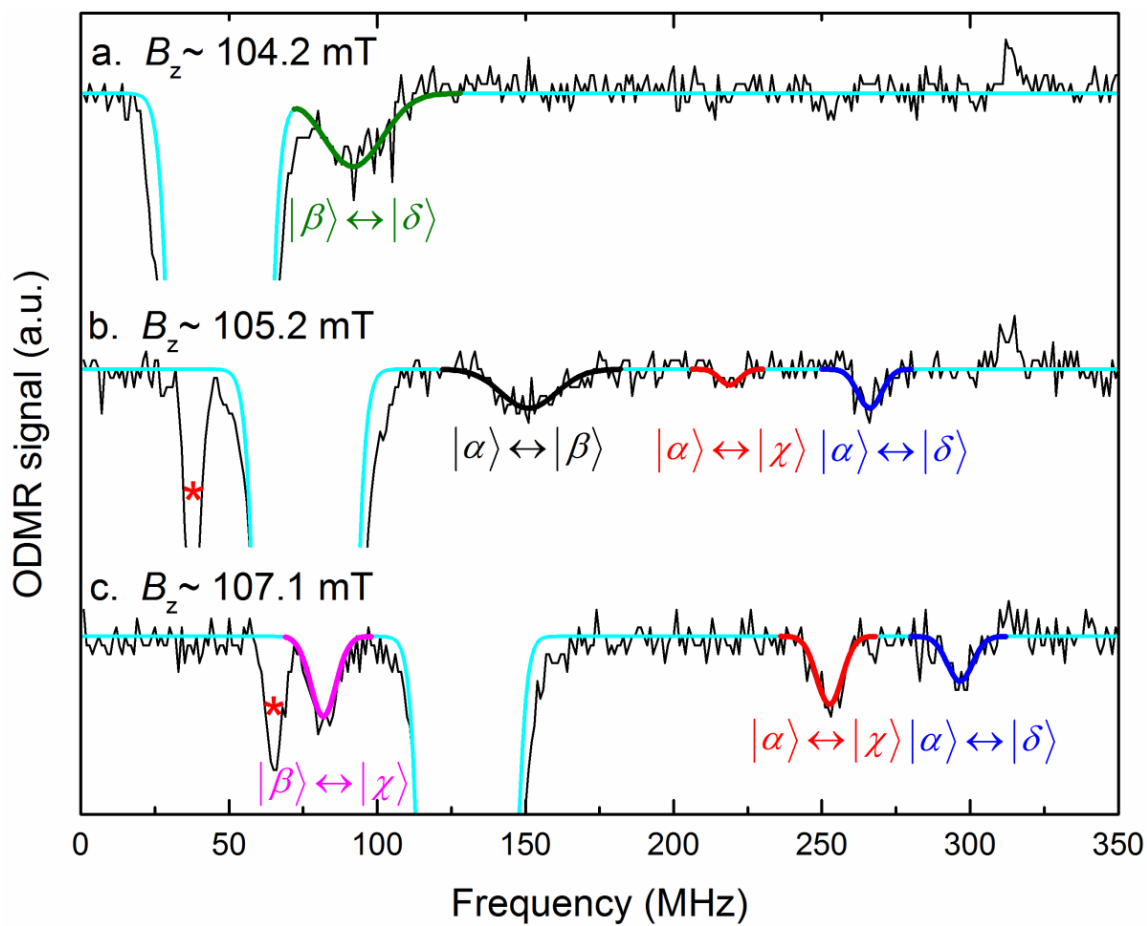

**Supplementary Figure S2 | Representative ODMR spectra near LAC.** The spectra are acquired at $B_z \sim 104.2$ mT (a), $B_z \sim 105.2$ mT (b), and $B_z \sim 107.1$ mT (c). All peaks are fit to Gaussian lineshapes. The intense primary peak associated with NV centres without $^{13}C_a$ is used to determine the magnetic field as shown in light blue solid lines. The weak peaks associated with NV-$^{13}C_a$ pair are colour-coded and labelled according to the related transitions determined in Fig. 4a. The peak labelled by a red * is an artefact whose second harmonic matches the frequency of the main peak.

**Supplementary Figure S3**

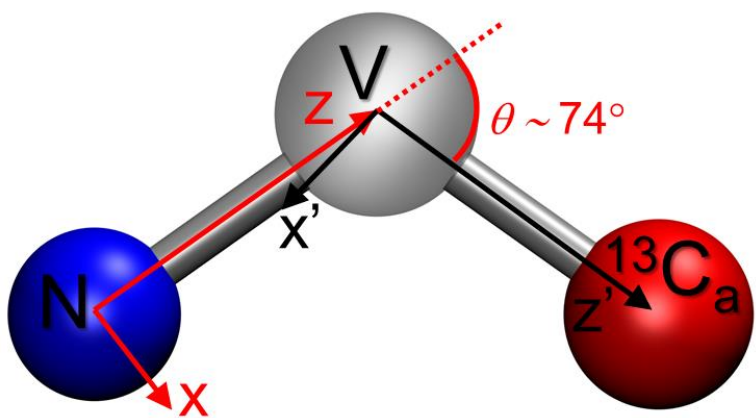

**Supplementary Figure S3 | Atomic representation of NV-$^{13}C_a$ pair.** The symmetry axis of the NV-$^{13}C_a$ hyperfine interaction is along the axis connecting the vacancy and $^{13}C_a$

**Supplementary Figure S4**

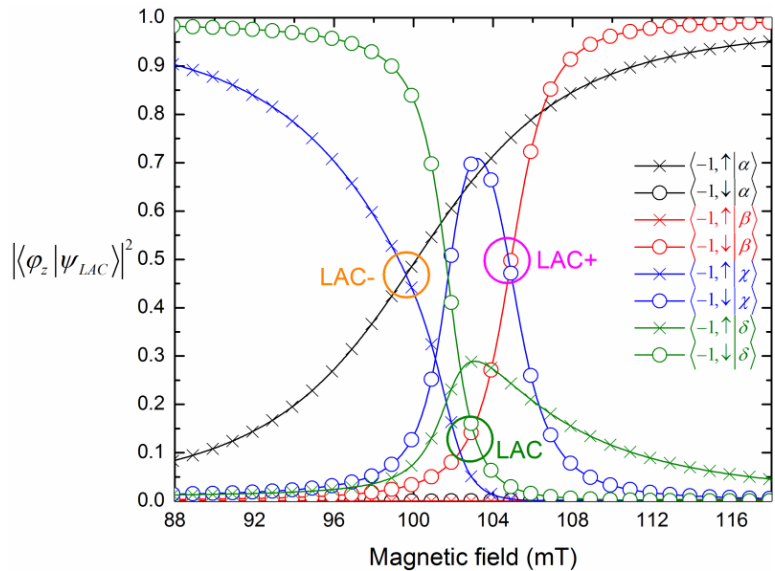

**Supplementary Figure S3 | Theoretical calculations of the eigenstates of NV-$^{13}C_a$ pair.** Eigenstates are expressed by their projections onto the $\left|m_S=-1, m_I=\uparrow/\downarrow\right\rangle_z$ states of the NV-$^{13}C_a$ pairs. The LAC-, LAC, and LAC+ are labelled by circles.

**Supplementary Figure S5**

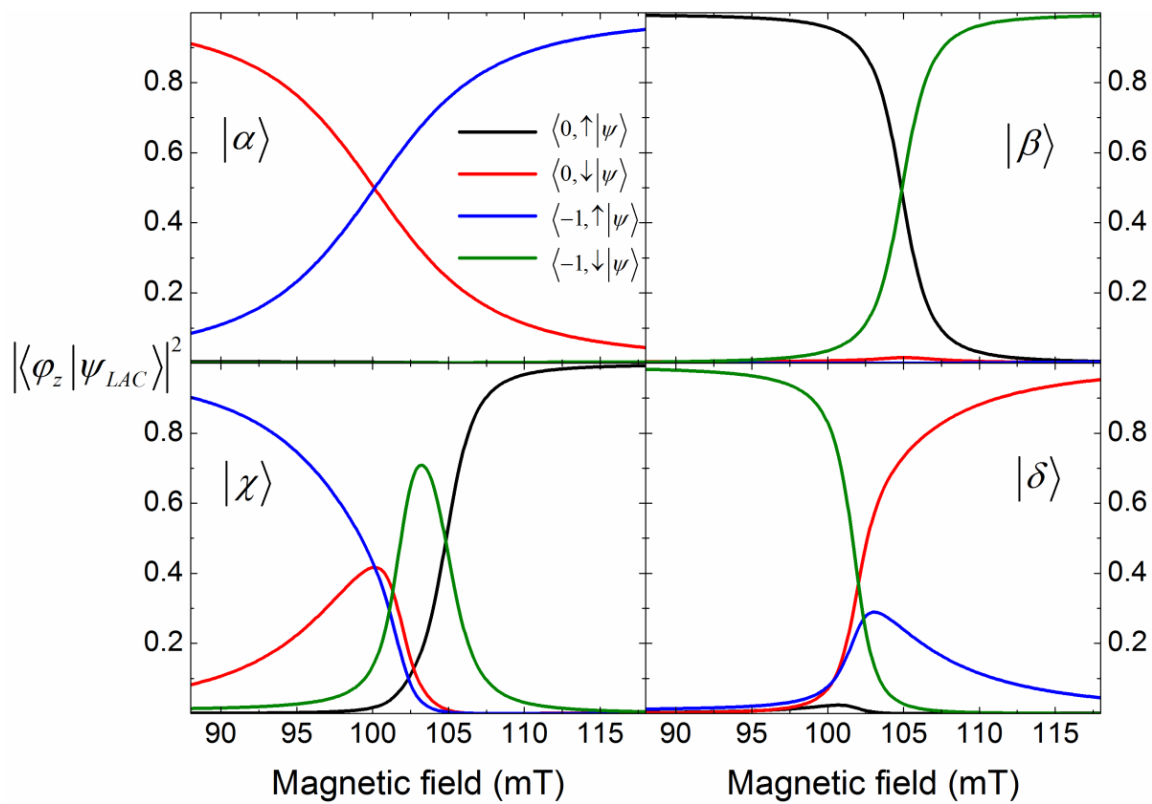

**Supplementary Figure S5 | Theoretical calculations of eigenstates in the NV-$^{13}$C$_a$ system.** Each mixed state is expressed by its respective projections onto the $\left| m_S = 0/-1, m_I = \uparrow / \downarrow \right\rangle_z$ states of the NV-$^{13}$C$_a$ pairs. This figure contains the same information as shown in Fig. 3b and Fig. S4.

**Supplementary Figure S6**

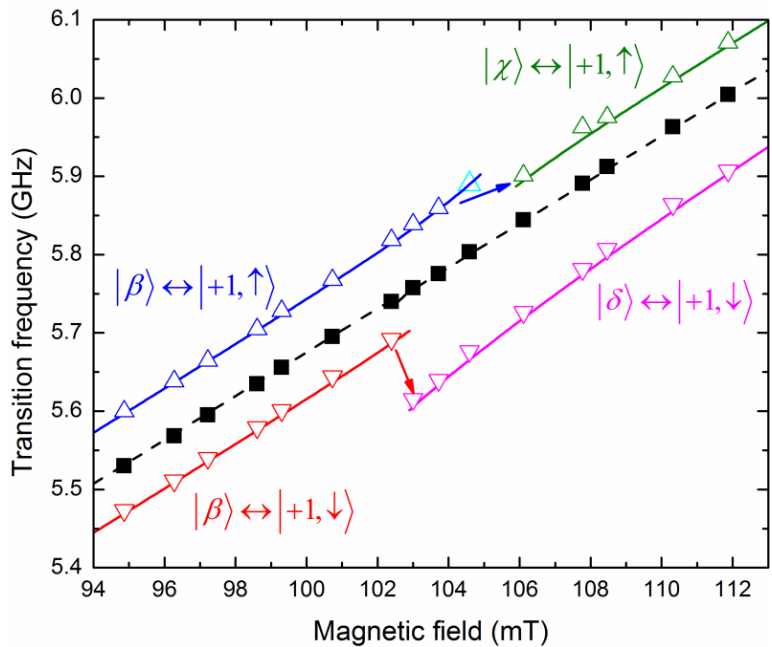

**Supplementary Figure S6 | Transition assignments for the estimation of nuclear polarisation.** The assignment of the observed transitions associated with the NV-$^{13}C_a$ pairs (open symbols) or with NV without $^{13}C_a$ (black solid squares), based on the measured frequency in comparison with theoretical calculation (solid lines for the NV-$^{13}C_a$ pairs, and dashed line for the NV without $^{13}C_a$). The disappearance of the $|\beta\rangle \leftrightarrow |+1,\uparrow\rangle$ transition near LAC+ is shown by an open, light blue up triangle.